\documentclass{aipproc}

\usepackage{amsmath}
\usepackage{amssymb}

\layoutstyle{6x9}

\SetInternalRegister\hbadness{8000} % pseudo latin isn't breaking very well :-)

\newcommand\doingARLO[2][]{%
  \ifx\mmref\undefined #1\else #2\fi
}

\begin{document}
\def\esp #1{e^{\displaystyle{#1}}}
\def\slash#1{\setbox0=\hbox{$#1$}#1\hskip-\wd0\dimen0=5pt\advance
       \dimen0 by-\ht0\advance\dimen0 by\dp0\lower0.5\dimen0\hbox
         to\wd0{\hss\sl/\/\hss}}\def\ink {\int~{d^4k\over (2\pi)^4}~}

\title
      [Nonperturbative calculations in SU(3) gauge theory]
      {Nonperturbative calculations in SU(3) gauge theory}

\classification{12.38.Aw, 12.38.Lg}
\keywords{confinement, nonperturbative quantization}

\author{Vladimir Dzhunushaliev}{
  address={Dept. Phys. and Microel. Eng., KRSU, Kievskaya Str. 44, 
  720021, Bishkek, Kyrgyz Republic.},
  email={dzhun@hotmail.kg},
  %thanks={This work was commissioned by the AIP}
}
\author{Douglas Singleton}{
  address={Physics Dept., CSU Fresno, M/S MH37 Fresno, CA 93740-8031, USA.},
  email={dougs@csufresno.edu},
}
\author{Tatyana Nikulicheva}{
  address={Dept. Phys. and Microel. Eng., KRSU, Kievskaya Str. 44, 
  720021, Bishkek, Kyrgyz Republic.}
}

%\begin{abstract}

%\end{abstract}

\date{\today}

\maketitle

Modern perturbative quantum field theories (QFT) are based on the concept of a
particle or field quanta. This perturbative paradigm gives the
most accurate description for some aspects of the physical world
({\it e.g.} the $g$-factor of the electron). 
Mathematically the elementary particles that arise in 
perturbative quantum field theory are quantized harmonic excitations
of fundamental  fields. The quanta are defined
through the creation and annihilation operators, $a^{\dagger}$ and
$a$, of the ``second-quantized" theory. 
\par
The physical reason of the appearance of quanta is that in linear
theories, such as electrodynamics, a general solution can be
obtained by making the Fourier expansion:
\begin{equation}
A_{\mu}^{linear} = \int d^3 k \sum_{\lambda = 0}^3 [a_k (\lambda)
\epsilon_{\mu}(k, \lambda) e^{-i k \cdot x} + a_k^{\dagger}
(\lambda) \epsilon_{\mu}^{\ast} (k, \lambda) e^{i k \cdot x}],
\label{intr-50}
\end{equation}
where $\epsilon_{\mu}$ is the polarization vector. The key point is 
that the general solution can be obtained by a linear superposition
of the plane wave solutions of the theory.
\par
For a theory based on a non-Abelian group, like quantum
chromodynamics (QCD), this can no longer be done in the strong
coupling limit, due to the nonlinear nature of Yang-Mills
equations. 
There are certain nonperturbative field configurations (for example, the
hypothesized color electric flux tube that is thought to be
important in the dual superconducting picture of confinement) in which the
field distribution can not be explained as a cloud of quanta. One way of looking at this
situation is that the fields of these nonperturbative configurations
are split into ordered fields (the fields inside flux tube stretched
between quark and antiquark) and disordered fields (the fields outside the
flux tube). These fields can not be interpreted as a (perturbative) cloud
of quanta. In such situations the fields play a primary role over the
particles.
\par
The need for nonperturbative techniques in strongly
interacting, nonlinear quantum field theories is an old
problem, and much effort has gone into trying to
find an appropriate frame in which to carry on calculations
for these theoreies. Despite this the problem is not yet fully resolved.
\par 
In \cite{dzhsin1} a possible approach was suggested based
on a version of a quantization method originally due to
Heisenberg. Starting with the classical SU(3) Yang-Mills equations 
\begin{equation}
    \partial_\nu \mathcal F^{B\mu\nu} = 0
\label{sec2-1-10}
\end{equation}
($\mathcal F^B_{\mu \nu} = \partial_\mu \mathcal A^B_\nu -
\partial_\nu \mathcal A^B_\mu + g f^{BCD} \mathcal A^C_\mu \mathcal A^D_\nu$
is the SU(3) field strength) one replaces the classical fields by field operators
$\mathcal A^B_{\mu} \rightarrow \widehat{\mathcal A}^B_\mu$. This yields the
following differential equations for the operators
\begin{equation}
    \partial_\nu \widehat {\mathcal F}^{B\mu\nu} = 0.
\label{sec2-1-20}
\end{equation}
These nonlinear equations for the field operators of
the nonlinear quantum fields can be used to determine
expectation values for the field operators
$\widehat {\mathcal A}^B_\mu$. One can also use these
equations to determine the expectation values of operators
that are built up from the fundamental operators
$\widehat {\mathcal A}^B_\mu$. 
The simple gauge field expectation values,
$\langle \mathcal{A}_\mu ^B (x) \rangle$, are obtained by
averaging Eq. \eqref{sec2-1-20} over some quantum state $| Q \rangle$
\begin{equation}
  \left\langle Q \left|
  \partial_\nu \widehat {\mathcal F}^{B\mu\nu}
  \right| Q \right\rangle = 0.
\label{sec2-1-30}
\end{equation}
One problem in using these equations to obtain expectation values
like $\langle \mathcal A^B_\mu \rangle$, is that these equations
involve not only powers or derivatives of $\langle \mathcal
A^B_\mu \rangle$ ({\it i.e.} terms like $\partial_\alpha \langle
\mathcal A^B_\mu \rangle$ or $\partial_\alpha
\partial_\beta \langle \mathcal A^B_\mu \rangle$) but also contain
terms like $\mathcal{G}^{BC}_{\mu\nu} = \langle \mathcal A^B_\mu
\mathcal A^C_\nu \rangle$. Starting with Eq. \eqref{sec2-1-30} one
can generate an operator differential equation for this product
$\widehat {\mathcal A}^B_\mu \widehat {\mathcal A}^C_\nu$
allowing the determination of the Green's function
$\mathcal{G}^{BC}_{\mu\nu}$
\begin{equation}
  \left\langle Q \left|
  \widehat {\mathcal A}^B(x) \partial_{y\nu} \widehat {\mathcal F}^{B\mu\nu}(y)
  \right| Q \right\rangle = 0.
\label{sec2-1-40}
\end{equation}
However this equation will in turn contain other, higher order
Green's functions. Repeating these steps leads to an infinite set
of equations connecting Green's functions of ever increasing
order. This construction, leading to an infinite set of coupled,
differential equations, does not have an exact, analytical solution
and so must be handled using some approximation.
\par
In order to do some calculations we give an approximate method which leads to the 2
and 4-points Green's functions only \cite{dzhsin1}. We will consider two cases:
in the first one the fields are in completely disordered phase. Starting with 
the pure SU(3) Lagrangian
\begin{equation}
\widehat {\mathcal L}_{SU(3)} = \frac{1}{4}
  \widehat {\mathcal F}^B_{\mu \nu}\widehat {\mathcal F}^{B \mu \nu}
\label{sec4-270}
\end{equation} 
one can arrive at following effective, pure scalar Lagrangian (see \cite{dzhsin1} for details)  
\begin{equation}
\begin{split}
    \frac{g^2}{4} 
    \mathcal {L}_{eff} = &- \frac{1}{2}\left( \partial_\mu \phi^A \right)^2 +
    \frac{\lambda_1}{4} \left[ \phi^a \phi^a - \phi^a_0 \phi^a_0
    \right]^2  +
    \frac{\lambda_2}{4}
    \left[ \phi^m \phi^m -  \phi^m_0 \phi^m_0
    \right]^2 + 
    \\
    &\left( \phi^a \phi^a \right) \left( \phi^m \phi^m \right) 
\label{sec4-275}
\end{split}
\end{equation}
where the vector gauge fields have been replaced by effective scalar field, $\phi ^A$,
through the ansatz
\begin{equation}
    \mathcal{G}^{BC}_{\alpha \beta} (x,y) \approx
    -\eta_{\alpha \beta} f^{BDE} f^{CDF} \phi^E (x) \phi^F(y)
\label{sec4-50}
\end{equation}
The $a$ index in eq. (\ref{sec4-275}) refers to the SU(2) subgroup of SU(3), and
the $m$ index refers to the coset SU(3)/SU(2), and two separate couplings 
($\lambda _1$ and $\lambda _2$) have been introduced between the SU(2) 
subgroup and the coset. A numerical investigation of eq. (\ref{sec4-275}) yields
a regular solution  with finite energy. The profile of the energy density for this solution is
given in the Fig. 1A. We interpret this solution as a glueball.
\par 
A second example involves both an ordered (described
by a vector gauge field) and disordered (described by an
effective scalar field) phase. Again after some assumptions and 
simplifications \cite{dzhsin1} we have find regular solutions with 
finite energy density.  The effective Lagrangian in this case
has the form 
\begin{equation}
\begin{split}
  \mathcal{L}_{eff} = - \frac{1}{4}  h^a_{\mu\nu} h^{a\mu\nu} +
  \frac{1}{2}   
  \left(D_\mu \phi^a\right)^2 
  - \frac{\lambda}{4} \left( \phi^a (x) \phi^a(x) \right)^2  + 
  \frac{g^2}{2} a_{\mu} ^b \phi ^b a^{c \mu} \phi ^c 
\label{sec5-470}
\end{split}
\end{equation}
where $h^b_{\mu\nu} = \partial_\mu a^b_\nu - \partial_\nu a^b_\mu + 
\epsilon^{bcd}a^c_\mu a^d_\nu$ 
is the SU(2) gauge field (ordered phase) and $\phi^a$ 
is a scalar field which describes 2 and 4-points Green's functions of 
SU(3)/SU(2) coset fields (disordered phase) via a relationship
similar to eq. (\ref{sec4-50}) {\it i.e.} $\mathcal{G}^{mn} (x,y) =
\frac{1}{3} f^{mpb} f^{npc} \phi^b (x) \phi^c(y)$ . A numerical investigation
of the effective Lagrangian of eq. (\ref{sec5-470}) shows that there are
finite energy solutions whose profiles for the longitudinal color electric field 
and transversal electric and magnetic fields are given on Fig. 1B.
\begin{figure}
  \begin{minipage}[t]{.45\linewidth}
    \fbox{
    \includegraphics[height=5cm,width=5cm]{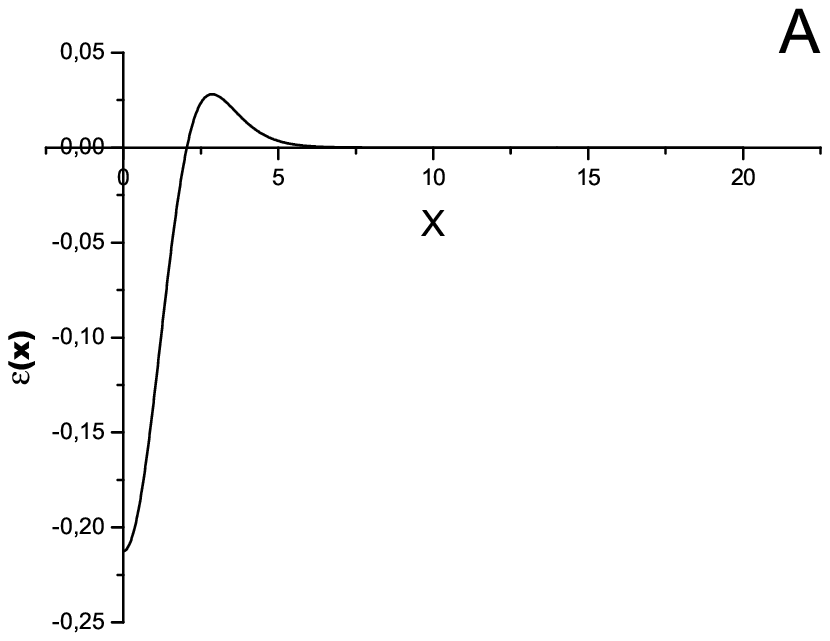}
    }
%    \caption{Fig. 1. The energy density of glueball/soliton solution. 
%    Fig.2. The color electric $E^3_z(x)$, $E^1_\rho(x)$ and magnetic
%    $H^2_\varphi (x)$ fields}
%    \label{fig:f-reg}
  \end{minipage}\hfill
  \begin{minipage}[t]{.45\linewidth}
    \fbox{
        \includegraphics[height=5cm,width=5cm]{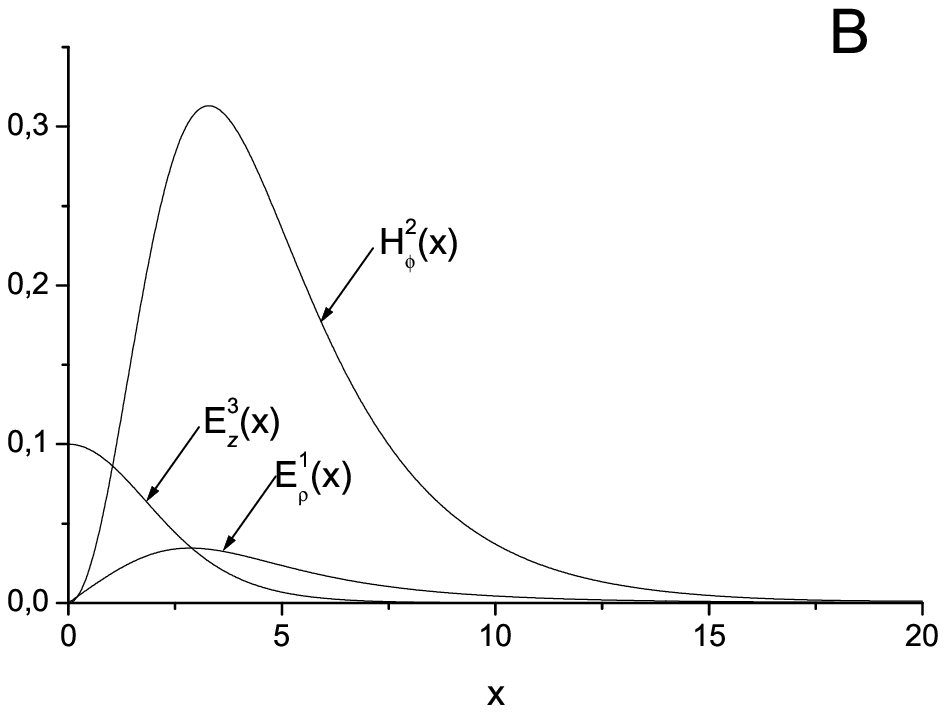}}
    \caption{\textbf{A}. The energy density of glueball/soliton solution. 
    \textbf{B}. The color electric $E^3_z(x)$, $E^1_\rho(x)$ and magnetic
    $H^2_\varphi (x)$ fields}
    \label{fig4}
  \end{minipage}\hfill
\end{figure}

In conclusion by applying a Heisenberg-like quantization method to
pure Yang-Mills theory we are able to construct effective Lagrangians that
have only scalar fields or a mixture of scalar plus gauge fields. Both of
these systems have finite energy solutions which are phenomenologically 
interesting. The completely disordered phase Lagrangian, containing only
effective scalar fields, had finite energy solutions which could be interpreted as
glueballs. The second case had a Lagrangian with both ordered (the SU(2)
gauge field) and disordered (the effective scalar field) phases. This system 
had finite energy flux tube-like solutions.

\end{document}